\documentstyle[12pt,epsfig]{article}
\input epsf

\begin{document}

%%%%% The following lines create the SLAC Pub Title Page
%%
\thispagestyle{empty}
\renewcommand{\thefootnote}{\fnsymbol{footnote}}

%%%%% Substitute your Pub number, month and year in the following:
%%
\begin{flushright}
{\small
SLAC--PUB--8553\\
August 2000\\}
\end{flushright}

\vspace{.8cm}

%%%%% Title and Author Information:
%%
\begin{center}
{\large \bf
Light Quark Fragmentation in Polarized $Z^{0}$ Decays at SLD
\footnote{Work supported in part by Department of Energy contract
  DE-AC03-76SF00515.
}}

\vspace{1cm}

M. Kalelkar\\
Rutgers University, Piscataway, NJ 08854\\
\vspace{1cm}

Representing The SLD Collaboration$^{**}$\\
Stanford Linear Accelerator Center, Stanford University,
Stanford, CA  94309\\

\end{center}

\vfill

\begin{center}
{\large
Abstract }
\end{center}

%===============
% Abstract
%===============
{
We report results on two physics topics from the SLD experiment at the SLAC
Linear Collider, using our full sample of 550,000 events of the type
$e^{+}e^{-} \rightarrow Z^{0} \rightarrow q\bar{q}$.  The electron beam was
polarized, enabling the quark and antiquark hemispheres to be tagged in each
event.  One physics topic is the first study of rapidities signed such that
positive rapidity is along the quark rather than antiquark direction. 
Distributions of ordered differences in signed rapidity between pairs of
particles are analyzed, providing the first direct observation of baryon number
ordering along the $q\bar{q}$ axis.  The other topic is the first direct
measurement of $A_{s}$, the parity-violating coupling of the $Z^{0}$ to strange
quarks, by measuring the left-right forward-backward production asymmetry in
polar angle of the tagged $s$ quark.  We obtain $A_{s}$ = $0.895\pm
0.066(stat.)\pm 0.062(syst.)$, which is consistent with the Standard Model and
is currently the most precise measurement of this quantity.
}
\vfill

%%%%%%%%%%%%%%%
%% Choose"Presented at," "Contributed to" for conference papers
%% or "Submitted to" for journal papers
%%%%%%%%%%%%%%%
\begin{center} 

{\it Presented at the International Euroconference on Quantum Chromodynamics
(QCD 00), 6-13 July 2000, Montpellier, France}
\end{center}

\newpage
%%
%%%%% End of title page

%%%%% Following are the commands to create the rest of the SLAC Pub.
%%
%%%%% The next two lines change the line spacing to doublespace,
%%      if you should need to do that.
%%
%\renewcommand{\baselinestretch}{2}
%\normalsize

%%%%% Your paper starts here:
%%

%% To get page numbers in the rest of the paper:
%
\pagestyle{plain}

\section{Introduction}

In this paper we report new results on two physics topics from the SLD
experiment at the SLAC Linear Collider, using our full data sample of 550,000
hadronic decays of $Z^{0}$ bosons produced in $e^{+}e^{-}$ annihilations.  A
unique feature of the experiment was a highly longitudinally polarized electron
beam, with an average magnitude of polarization of 73\%.

One physics topic is a study of signed rapidities.  The rapidity of a particle
is typically defined with an arbitrary sign.  If a sign could be given to each
measured rapidity such that, for example, positive (negative) rapidity
corresponds to the initial quark (antiquark) direction, then one could measure
the extent to which a leading particle has higher rapidity than its associated
antiparticle, and the extent to which low-momentum particles in jets remember
the initial quark/antiquark direction.

The other physics topic in this paper is the first direct measurement of the
strange quark asymmetry parameter $A_{s}$.  Measurements of the fermion
production asymmetries in the process $e^{+}e^{-} \rightarrow Z^{0} \rightarrow
f\bar{f}$ provide information on the extent of parity violation in the coupling
of the $Z^{0}$ bosons to fermions of type $f$.  The differential production
cross section can be expressed in terms of $x = \cos\theta$, where $\theta$ is
the polar angle of the final state fermion $f$ with respect to the electron
beam direction: $${d\sigma\over{dx}} \propto (1-A_{e}P_{e})(1+x^{2}) +
2A_{f}(A_{e}-P_{e})x$$ where $P_{e}$ is the longitudinal polarization of the
electron beam, the positron beam is assumed unpolarized, and the asymmetry
parameters $A_{f} = 2v_{f}a_{f}/(v_{f}^{2} + a_{f}^{2})$ are defined in terms
of the vector and axial-vector couplings of the $Z^{0}$ to fermion $f$.  The
Standard Model predictions for the values of the asymmetry parameters,
assuming $\sin^{2}\theta_{w} = 0.23$, are $A_{e} = A_{\mu} = A_{\tau} = 0.16$,
$A_{u} = A_{c} = A_{t} = 0.67$, and $A_{d} = A_{s} = A_{b} = 0.94$.  For a
given final state $f\bar{f}$, if one measures the polar angle distributions in
equal luminosity samples taken with negative and positive beam polarization,
then one can derive the left-right forward-backward asymmetry:
$$\tilde{A}^{f}_{FB} = {3\over{4}}\mid P_{e}\mid A_{f}$$ which is insensitive
to the initial state coupling.

A number of previous measurements have been made of the leptonic asymmetries
and the heavy-flavor asymmetries, but very few measurements exist for the light
quark flavors, due to the difficulty of tagging specific light flavors. We
present a direct measurement of the strange quark asymmetry parameter $A_{s}$,
in which $Z^{0} \rightarrow s\bar{s}$ events were tagged by the absence of $B$
or $D$ hadrons and the presence in each hemisphere of a high-momentum $K^{\pm}$
or $K^{0}_{S}$.

\section{Particle Identification}

A description of the SLD detector, trigger, track and hadronic event selection,
and Monte Carlo simulation is given in Ref.~\cite{impact}. The identification
of $\pi^{\pm}$, K$^{\pm}$, p, and $\bar{\rm p}$ was achieved by reconstructing
emission angles of individual Cherenkov photons radiated by charged particles
passing through liquid and gas radiator systems of the SLD Cherenkov Ring
Imaging Detector (CRID)~\cite{crid}.  In each momentum bin, identified $\pi$,
K, and p were counted, and these were unfolded using the inverse of an
identification efficiency matrix~\cite{pavel}, and corrected for track
reconstruction efficiency.  The elements of the identification efficiency
matrix were mostly measured from data, using selected $K^{0}_{S}$, $\tau$, and
$\Lambda$ decays.  A detailed Monte Carlo simulation was used to derive the
unmeasured elements in terms of these measured ones.

$K^{0}_{S}\rightarrow\pi^{+}\pi^{-}$ and $\Lambda^{0} (\bar{\Lambda}^{0})
\rightarrow p\pi^{-} (\bar{p}\pi^{+})$  decays were reconstructed as described
in Ref.~\cite{bfp,staengle} by examining appropriate invariant mass
distributions.

\section{Signed Rapidities}

We next tagged the quark (vs.\ antiquark) direction in each hadronic event by
using the electron beam polarization for that event, exploiting the large
forward-backward quark production asymmetry in $Z^{0}$ decays.  If the beam was
left(right)-handed, then the thrust axis was signed such that $\cos\theta_{T}$
was positive (negative).  Events with $|\cos\theta_{T}| < 0.15$ were removed,
as the production asymmetry is small in this region.  The probability to tag
the quark direction correctly in these events was 73\%, assuming Standard Model
couplings at tree-level.

\begin{figure}[htbp]
\centering
\epsfxsize=15.0cm
\leavevmode
\epsfbox{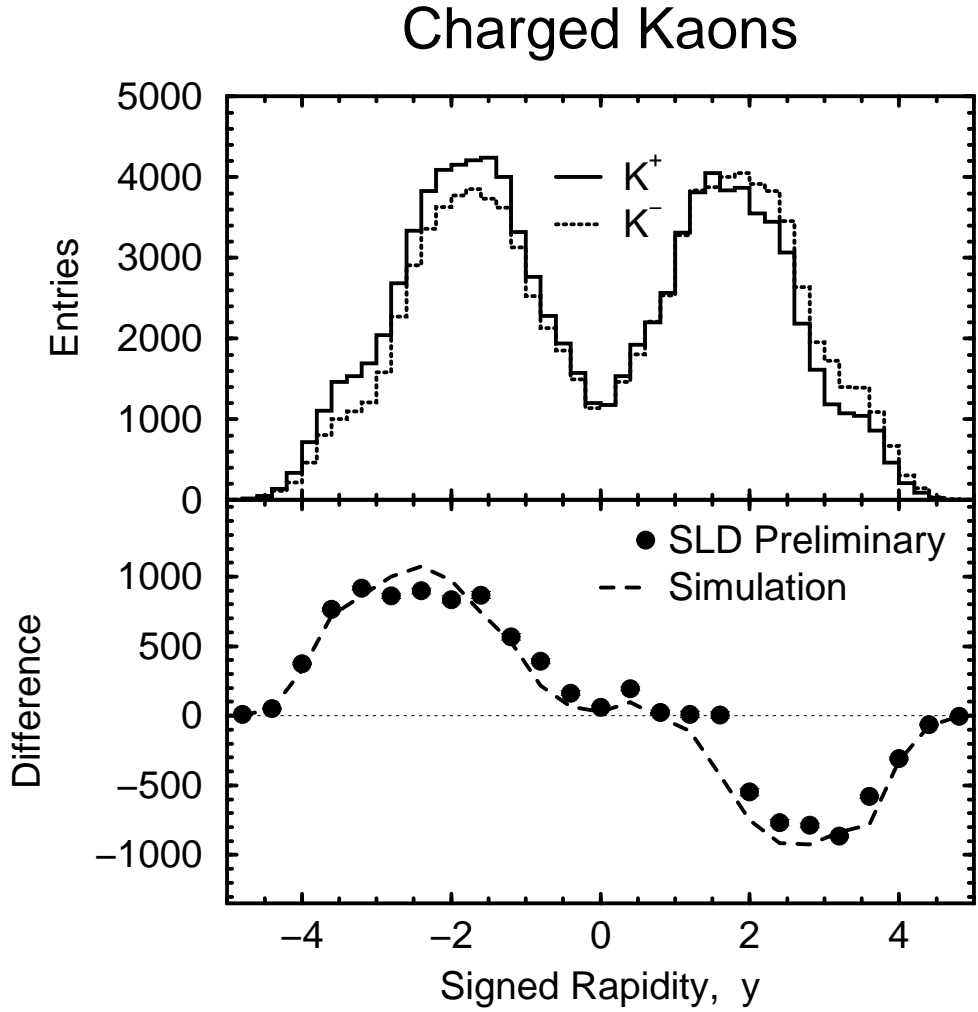}
\caption{\label{fig1}
Top: Distributions of the rapidity with respect to the signed thrust axis for
positively (solid histogram) and negatively (dashed) charged kaons.  Bottom:
The difference between these two distributions compared with the prediction of
the Monte Carlo simulation.}
\end{figure}

For each identified particle the rapidity $y =$ 0.5 ln$((E+p_{\|})/(E-p_{\|}))$
was calculated using the measured momentum and its projection $p_{\|}$ along
the thrust axis, and the appropriate hadron mass.  The rapidity with respect to
the signed thrust axis is naturally signed such that positive rapidity
corresponds to the hemisphere in the tagged direction of the initial quark, and
negative rapidity corresponds to the tagged antiquark hemisphere.  The signed
rapidity distributions for identified $K^{+}$ and $K^{-}$ are shown in
Fig.~\ref{fig1}. 
 
There is a clear difference between the two, with more $K^{-}$ than  $K^{+}$ in
the quark hemisphere, as expected due to leading  $K^{-}$  produced in
$s$-quark jets~\cite{bfp,leading}.  The difference between the two
distributions is also shown in the figure and is compared with the prediction
of the JETSET~\cite{jetset} simulation, which is consistent with the data.

\begin{figure}[htbp]
\centering
\epsfxsize=15.0cm
\leavevmode
\epsfbox{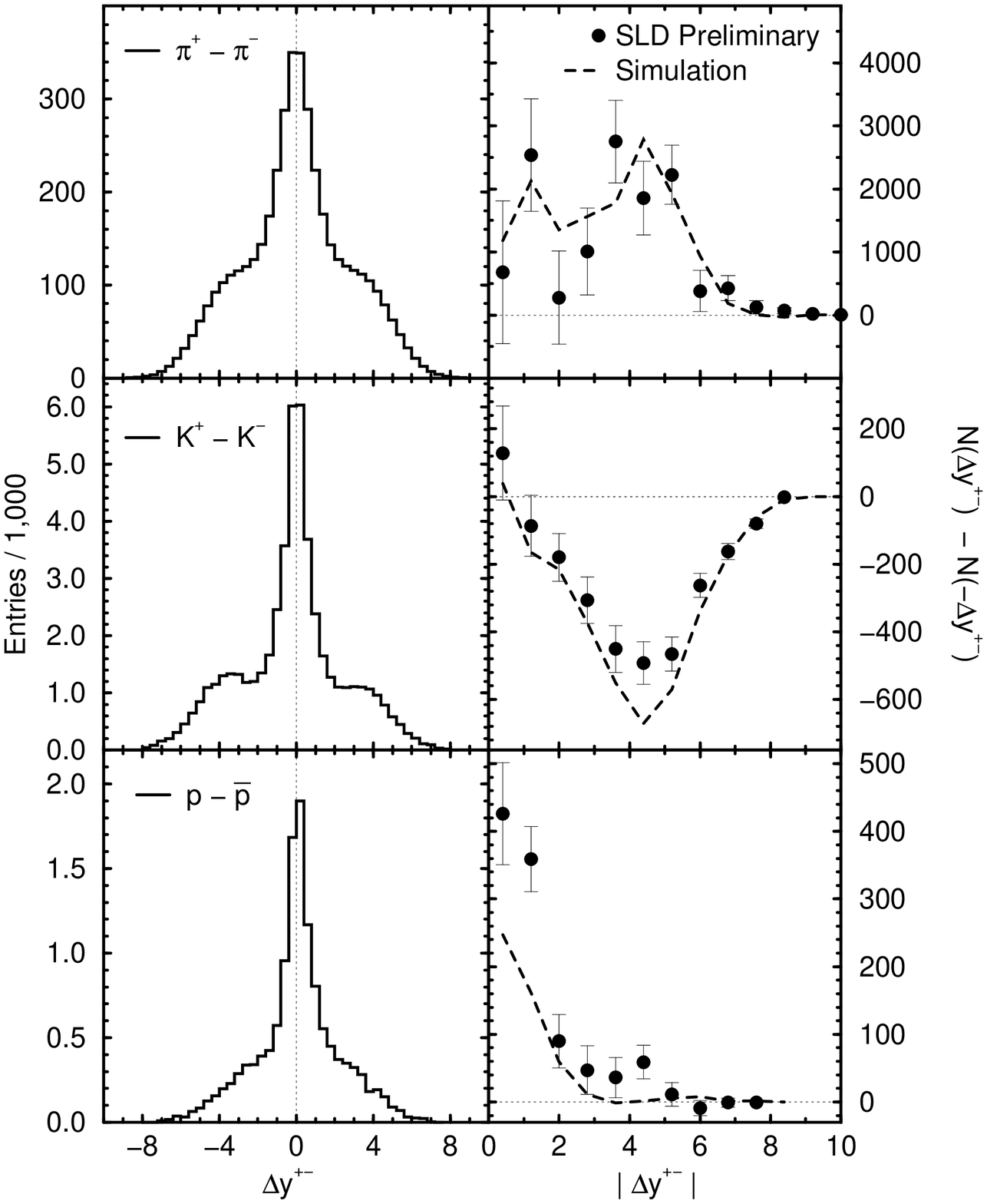}
\caption{\label{fig2}
Left: Distributions of the difference between the signed rapidities of
positively and negatively charged hadrons of the same type.  Right: Differences
between the right and left sides of the distributions, compared with the
prediction of the Monte Carlo simulation.}
\end{figure}

For pairs of identified particles, one can define an ordered rapidity
difference.  For particle-antiparticle pairs, we define $\Delta y^{+-} = y_{+}
- y_{-}$ as the difference between the signed rapidities of the positively
charged particle and the negatively charged particle.  In Fig.~\ref{fig2} we
show the distribution of $\Delta y^{+-}$ for $\pi^{+}\pi^{-}$, $K^{+}K^{-}$ and
$p\bar{p}$ pairs.
Asymmetries in these distributions are indications of
ordering along the event axis, and the differences between the positive and
negative sides of these distributions are also shown.  The predictions of the
simulation are also shown and are consistent with the data at high $|\Delta
y^{+-}|$.

The negative difference at high $|\Delta y^{+-}|$ for the $K^{+}K^{-}$ pairs
can be attributed to the fact that leading kaons are produced predominantly in
$s\bar{s}$ events.  For $\pi^{+}\pi^{-}$ pairs we observe a large positive
difference at high $|\Delta y^{+-}|$ rather than the expected small negative
difference, and we have confirmed that this effect is due entirely to
$c\bar{c}$ events.  Our sample of $uds$-tagged events does show the expected
small negative difference.

The positive difference in the lowest $|\Delta y^{+-}|$ bins for the $p\bar{p}$
pairs indicates that the baryon in an associated baryon-antibaryon pair follows
the quark direction more closely than the antibaryon.  This could be due to
leading baryon production and/or to baryon numbering ordering along the
entire fragmentation chain.  We find a significant effect in all of our
momentum bins, and the bulk of the observed difference occurs at low momentum. 
We therefore conclude that both of these effects contribute; this is the first
direct observation of baryon number ordering along the entire chain.  The
prediction of the simulation is low by a factor of two at low $|\Delta
y^{+-}|$.

\section{Strange Quark Asymmetry}

For the measurement of the strange quark asymmetry parameter $A_{s}$, the first
step was to select $s\bar{s}$ events and tag the $s$ and $\bar{s}$ jets.  We
used the SLD Vertex Detector~\cite{vxd} to measure each track's impact
parameter $d$ in the plane perpendicular to the beam direction.  We then
removed $c\bar{c}$ and $b\bar{b}$ events by requiring no more than one
well-measured track with $d$ larger than 2.5 times its uncertainty.

Each remaining event was divided into two hemispheres by the plane
perpendicular to the thrust axis.  In each hemisphere we searched for the
strange particle with the highest momentum.  If it was a charged kaon, we
required it to have $p>9$ GeV/c, while if it was a $K^{0}_{S}$ or
$\Lambda^{0}/\bar{\Lambda}^{0}$ it was required to have $p>5$ GeV/c.  An event
was tagged as $s\bar{s}$ if one hemisphere contained a $K^{\pm}$ selected as
just described, and the other contained either an oppositely charged $K^{\pm}$
or a $K^{0}_{S}$.  The thrust axis, signed so as to point into the hemisphere
containing (opposite) the $K^{-}(K^{+})$, was used as an estimate of the
initial $s$-quark direction.

\begin{figure}[htbp]
\centering
\epsfxsize=15.0cm
\leavevmode
\epsfbox{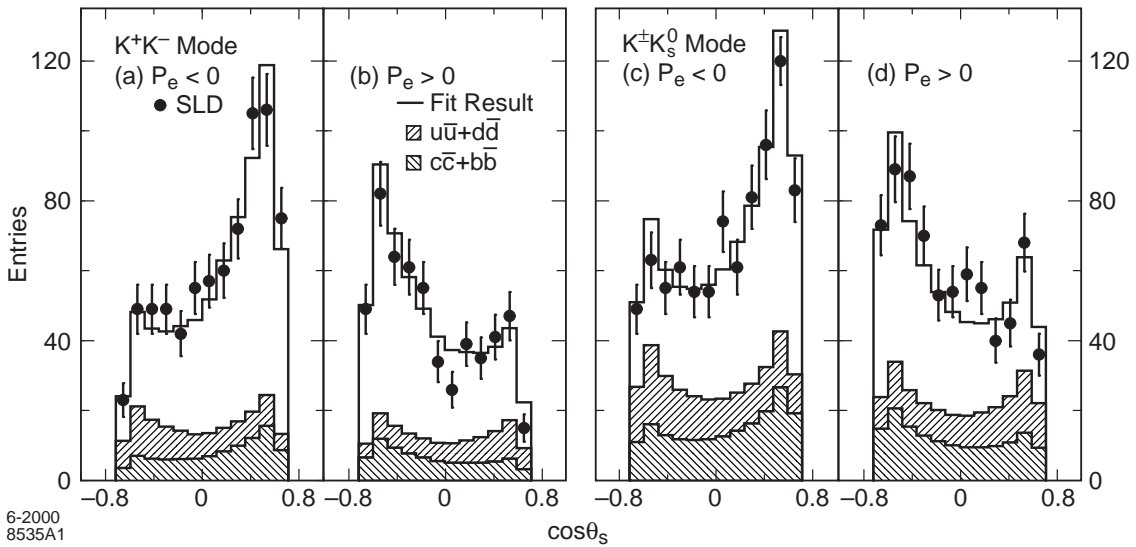}
\caption{\label{fig3}
Measured $s$-quark polar angle distributions (dots) for events in the (a,b)
$K^{+}K^{-}$ and (c,d) $K^{\pm}K^{0}_{S}$ modes, produced with (a,c) left- and
(b,d) right-polarized electron beams.  The histograms represent the result of a
simultaneous fit to the four distributions, and the upper (lower) hatched areas
indicate the estimated $u\bar{u}+d\bar{d}\ (c\bar{c}+b\bar{b})$ backgrounds.}
\end{figure}

Fig.~\ref{fig3} shows the distributions of the measured $s$-quark polar angle
$\theta_{s}$ for the $K^{+}K^{-}$ and $K^{\pm}K^{0}_{S}$ modes.  
In each case,
production asymmetries of opposite sign and different magnitude for left- and
right-polarized $e^{-}$ beams are visible.

$A_{s}$ was extracted from these distributions by a simultaneous maximum
likelihood fit, the result of which is shown as a histogram in the figure.  The
fit quality was good, with a $\chi^{2}$ of 42 for 48 bins.  Also shown in the
figure are our estimates~\cite{staengle} of the non-$s\bar{s}$ backgrounds,
which were mostly determined from data.  Our final result is $A_{s} = 0.895 \pm
0.066(stat.) \pm 0.062(syst.)$.  This result is consistent with the Standard
Model expectation of 0.935 for $A_{s}$, and with less precise previous
measurements~\cite{delphi,opal}.

\section{Conclusions}

We have presented results from a sample of 550,000 $e^{+}e^{-} \rightarrow
Z^{0} \rightarrow q\bar{q}$ events produced with a longitudinally polarized
electron beam.  Polarization enables us to give a sign to rapidities so that
positive rapidity corresponds to the quark (rather than antiquark) direction. 
The distribution of the difference between the signed rapidities of $K^{+}$ and
$K^{-}$ shows a large asymmetry at large values of the absolute rapidity
difference, a direct indication that the long-range correlated $KK$ pairs are
dominated by $s\bar{s}$ events.  There is a large asymmetry at small rapidity
difference for $p\bar{p}$ pairs, a clear indication of ordering of baryons
along the event axis.

We have also performed a measurement of $A_{s}$, the parity-violating coupling
of the $Z^{0}$ to strange quarks, obtained directly from the left-right
forward-backward production asymmetry in polar angle of the tagged $s$ quark. 
Our result is $A_{s}$ = $0.895\pm 0.066(stat.)\pm 0.062(syst.)$, which is
consistent with the Standard Model expectation, and with less precise previous
measurements.

\section{Questions and Answers}

Question (from M.\ Boutemeur, Munich): How do the $\pi^{\pm}$, $K^{\pm}$, and
$p/\bar{p}$ rates compare in quark and gluon jets?

Answer: My graduate student Hyejoo Kang is working on this very topic for her
Ph.D.\ thesis.  We expect to have results ready in time for the Moriond
Conference in early 2001.

\newpage
\section*{$^{**}$List of Authors} 
%
% author list for inclusion in LaTeX documents
% using \author{} and \address{} commands
%
% Instion number definitions:
%
\begin{center}
\def\iAOMORI{$^{(1)}$}
\def\iBRI{$^{(2)}$}
\def\iBRUN{$^{(3)}$}
\def\iBU{$^{(4)}$}
\def\iCOLO{$^{(5)}$}
\def\iCSU{$^{(6)}$}
\def\iFERR{$^{(7)}$}
\def\iFRAS{$^{(8)}$}
\def\iJHU{$^{(9)}$}
\def\iLBL{$^{(10)}$}
\def\iMASS{$^{(11)}$}
\def\iMISSI{$^{(12)}$}
\def\iMIT{$^{(13)}$}
\def\iMOSCOW{$^{(14)}$}
\def\iNAGO{$^{(15)}$}
\def\iOREG{$^{(16)}$}
\def\iOXF{$^{(17)}$}
\def\iPERU{$^{(18)}$}
\def\iRAL{$^{(19)}$}
\def\iRUTG{$^{(20)}$}
\def\iSLAC{$^{(21)}$}
\def\iSOONG{$^{(22)}$}
\def\iTENN{$^{(23)}$}
\def\iTOHO{$^{(24)}$}
\def\iUCSB{$^{(25)}$}
\def\iUCSC{$^{(26)}$}
\def\iVAND{$^{(27)}$}
\def\iWASH{$^{(28)}$}
\def\iWISC{$^{(29)}$}
\def\iYALE{$^{(30)}$}

  \baselineskip=.75\baselineskip
\mbox{Koya Abe\unskip,\iTOHO}
\mbox{Kenji Abe\unskip,\iNAGO}
\mbox{T. Abe\unskip,\iSLAC}
\mbox{I. Adam\unskip,\iSLAC}
\mbox{H. Akimoto\unskip,\iSLAC}
\mbox{D. Aston\unskip,\iSLAC}
\mbox{K.G. Baird\unskip,\iMASS}
\mbox{C. Baltay\unskip,\iYALE}
\mbox{H.R. Band\unskip,\iWISC}
\mbox{T.L. Barklow\unskip,\iSLAC}
\mbox{J.M. Bauer\unskip,\iMISSI}
\mbox{G. Bellodi\unskip,\iOXF}
\mbox{R. Berger\unskip,\iSLAC}
\mbox{G. Blaylock\unskip,\iMASS}
\mbox{J.R. Bogart\unskip,\iSLAC}
\mbox{G.R. Bower\unskip,\iSLAC}
\mbox{J.E. Brau\unskip,\iOREG}
\mbox{M. Breidenbach\unskip,\iSLAC}
\mbox{W.M. Bugg\unskip,\iTENN}
\mbox{D. Burke\unskip,\iSLAC}
\mbox{T.H. Burnett\unskip,\iWASH}
\mbox{P.N. Burrows\unskip,\iOXF}
\mbox{A. Calcaterra\unskip,\iFRAS}
\mbox{R. Cassell\unskip,\iSLAC}
\mbox{A. Chou\unskip,\iSLAC}
\mbox{H.O. Cohn\unskip,\iTENN}
\mbox{J.A. Coller\unskip,\iBU}
\mbox{M.R. Convery\unskip,\iSLAC}
\mbox{V. Cook\unskip,\iWASH}
\mbox{R.F. Cowan\unskip,\iMIT}
\mbox{G. Crawford\unskip,\iSLAC}
\mbox{C.J.S. Damerell\unskip,\iRAL}
\mbox{M. Daoudi\unskip,\iSLAC}
\mbox{S. Dasu\unskip,\iWISC}
\mbox{N. de Groot\unskip,\iBRI}
\mbox{R. de Sangro\unskip,\iFRAS}
\mbox{D.N. Dong\unskip,\iMIT}
\mbox{M. Doser\unskip,\iSLAC}
\mbox{R. Dubois\unskip,}
\mbox{I. Erofeeva\unskip,\iMOSCOW}
\mbox{V. Eschenburg\unskip,\iMISSI}
\mbox{E. Etzion\unskip,\iWISC}
\mbox{S. Fahey\unskip,\iCOLO}
\mbox{D. Falciai\unskip,\iFRAS}
\mbox{J.P. Fernandez\unskip,\iUCSC}
\mbox{K. Flood\unskip,\iMASS}
\mbox{R. Frey\unskip,\iOREG}
\mbox{E.L. Hart\unskip,\iTENN}
\mbox{K. Hasuko\unskip,\iTOHO}
\mbox{S.S. Hertzbach\unskip,\iMASS}
\mbox{M.E. Huffer\unskip,\iSLAC}
\mbox{X. Huynh\unskip,\iSLAC}
\mbox{M. Iwasaki\unskip,\iOREG}
\mbox{D.J. Jackson\unskip,\iRAL}
\mbox{P. Jacques\unskip,\iRUTG}
\mbox{J.A. Jaros\unskip,\iSLAC}
\mbox{Z.Y. Jiang\unskip,\iSLAC}
\mbox{A.S. Johnson\unskip,\iSLAC}
\mbox{J.R. Johnson\unskip,\iWISC}
\mbox{R. Kajikawa\unskip,\iNAGO}
\mbox{M. Kalelkar\unskip,\iRUTG}
\mbox{H.J. Kang\unskip,\iRUTG}
\mbox{R.R. Kofler\unskip,\iMASS}
\mbox{R.S. Kroeger\unskip,\iMISSI}
\mbox{M. Langston\unskip,\iOREG}
\mbox{D.W.G. Leith\unskip,\iSLAC}
\mbox{V. Lia\unskip,\iMIT}
\mbox{C. Lin\unskip,\iMASS}
\mbox{G. Mancinelli\unskip,\iRUTG}
\mbox{S. Manly\unskip,\iYALE}
\mbox{G. Mantovani\unskip,\iPERU}
\mbox{T.W. Markiewicz\unskip,\iSLAC}
\mbox{T. Maruyama\unskip,\iSLAC}
\mbox{A.K. McKemey\unskip,\iBRUN}
\mbox{R. Messner\unskip,\iSLAC}
\mbox{K.C. Moffeit\unskip,\iSLAC}
\mbox{T.B. Moore\unskip,\iYALE}
\mbox{M. Morii\unskip,\iSLAC}
\mbox{D. Muller\unskip,\iSLAC}
\mbox{V. Murzin\unskip,\iMOSCOW}
\mbox{S. Narita\unskip,\iTOHO}
\mbox{U. Nauenberg\unskip,\iCOLO}
\mbox{H. Neal\unskip,\iYALE}
\mbox{G. Nesom\unskip,\iOXF}
\mbox{N. Oishi\unskip,\iNAGO}
\mbox{D. Onoprienko\unskip,\iTENN}
\mbox{L.S. Osborne\unskip,\iMIT}
\mbox{R.S. Panvini\unskip,\iVAND}
\mbox{C.H. Park\unskip,\iSOONG}
\mbox{I. Peruzzi\unskip,\iFRAS}
\mbox{M. Piccolo\unskip,\iFRAS}
\mbox{L. Piemontese\unskip,\iFERR}
\mbox{R.J. Plano\unskip,\iRUTG}
\mbox{R. Prepost\unskip,\iWISC}
\mbox{C.Y. Prescott\unskip,\iSLAC}
\mbox{B.N. Ratcliff\unskip,\iSLAC}
\mbox{J. Reidy\unskip,\iMISSI}
\mbox{P.L. Reinertsen\unskip,\iUCSC}
\mbox{L.S. Rochester\unskip,\iSLAC}
\mbox{P.C. Rowson\unskip,\iSLAC}
\mbox{J.J. Russell\unskip,\iSLAC}
\mbox{O.H. Saxton\unskip,\iSLAC}
\mbox{T. Schalk\unskip,\iUCSC}
\mbox{B.A. Schumm\unskip,\iUCSC}
\mbox{J. Schwiening\unskip,\iSLAC}
\mbox{V.V. Serbo\unskip,\iSLAC}
\mbox{G. Shapiro\unskip,\iLBL}
\mbox{N.B. Sinev\unskip,\iOREG}
\mbox{J.A. Snyder\unskip,\iYALE}
\mbox{H. Staengle\unskip,\iCSU}
\mbox{A. Stahl\unskip,\iSLAC}
\mbox{P. Stamer\unskip,\iRUTG}
\mbox{H. Steiner\unskip,\iLBL}
\mbox{D. Su\unskip,\iSLAC}
\mbox{F. Suekane\unskip,\iTOHO}
\mbox{A. Sugiyama\unskip,\iNAGO}
\mbox{A. Suzuki\unskip,\iNAGO}
\mbox{M. Swartz\unskip,\iJHU}
\mbox{F.E. Taylor\unskip,\iMIT}
\mbox{J. Thom\unskip,\iSLAC}
\mbox{E. Torrence\unskip,\iMIT}
\mbox{T. Usher\unskip,\iSLAC}
\mbox{J. Va'vra\unskip,\iSLAC}
\mbox{R. Verdier\unskip,\iMIT}
\mbox{D.L. Wagner\unskip,\iCOLO}
\mbox{A.P. Waite\unskip,\iSLAC}
\mbox{S. Walston\unskip,\iOREG}
\mbox{A.W. Weidemann\unskip,\iTENN}
\mbox{E.R. Weiss\unskip,\iWASH}
\mbox{J.S. Whitaker\unskip,\iBU}
\mbox{S.H. Williams\unskip,\iSLAC}
\mbox{S. Willocq\unskip,\iMASS}
\mbox{R.J. Wilson\unskip,\iCSU}
\mbox{W.J. Wisniewski\unskip,\iSLAC}
\mbox{J.L. Wittlin\unskip,\iMASS}
\mbox{M. Woods\unskip,\iSLAC}
\mbox{T.R. Wright\unskip,\iWISC}
\mbox{R.K. Yamamoto\unskip,\iMIT}
\mbox{J. Yashima\unskip,\iTOHO}
\mbox{S.J. Yellin\unskip,\iUCSB}
\mbox{C.C. Young\unskip,\iSLAC}
\mbox{H. Yuta\unskip.\iAOMORI}

\it
  \vskip \baselineskip                   % \bigskip did not work
%  \centerline{(The SLD Collaboration)}   % include collaboration name
%  \vskip \baselineskip
  \baselineskip=.75\baselineskip   % shrink the interline spacing
\iAOMORI
  Aomori University, Aomori, 030 Japan, \break
\iBRI
  University of Bristol, Bristol, United Kingdom, \break
\iBRUN
  Brunel University, Uxbridge, Middlesex, UB8 3PH United Kingdom, \break
\iBU
  Boston University, Boston, Massachusetts 02215, \break
\iCOLO
  University of Colorado, Boulder, Colorado 80309, \break
\iCSU
  Colorado State University, Ft. Collins, Colorado 80523, \break
\iFERR
  INFN Sezione di Ferrara and Universita di Ferrara, I-44100 Ferrara, Italy,
\break
\iFRAS
  INFN Laboratori Nazionali di Frascati, I-00044 Frascati, Italy, \break
\iJHU
  Johns Hopkins University,  Baltimore, Maryland 21218-2686, \break
\iLBL
  Lawrence Berkeley Laboratory, University of California, Berkeley, California
94720, \break
\iMASS
  University of Massachusetts, Amherst, Massachusetts 01003, \break
\iMISSI
  University of Mississippi, University, Mississippi 38677, \break
\iMIT
  Massachusetts Institute of Technology, Cambridge, Massachusetts 02139, \break
\iMOSCOW
  Institute of Nuclear Physics, Moscow State University, 119899 Moscow, Russia,
\break
\iNAGO
  Nagoya University, Chikusa-ku, Nagoya, 464 Japan, \break
\iOREG
  University of Oregon, Eugene, Oregon 97403, \break
\iOXF
  Oxford University, Oxford, OX1 3RH, United Kingdom, \break
\iPERU
  INFN Sezione di Perugia and Universita di Perugia, I-06100 Perugia, Italy,
\break
\iRAL
  Rutherford Appleton Laboratory, Chilton, Didcot, Oxon OX11 0QX United Kingdom,
\break
\iRUTG
  Rutgers University, Piscataway, New Jersey 08855, \break
\iSLAC
  Stanford Linear Accelerator Center, Stanford University, Stanford, California
94309, \break
\iSOONG
  Soongsil University, Seoul, Korea 156-743, \break
\iTENN
  University of Tennessee, Knoxville, Tennessee 37996, \break
\iTOHO
  Tohoku University, Sendai, 980 Japan, \break
\iUCSB
  University of California at Santa Barbara, Santa Barbara, California 93106,
\break
\iUCSC
  University of California at Santa Cruz, Santa Cruz, California 95064, \break
\iVAND
  Vanderbilt University, Nashville,Tennessee 37235, \break
\iWASH
  University of Washington, Seattle, Washington 98105, \break
\iWISC
  University of Wisconsin, Madison,Wisconsin 53706, \break
\iYALE
  Yale University, New Haven, Connecticut 06511. \break

\rm
%
%  }   % end of address list

\end{center}

\end{document}